\documentclass[prl,aps,floats,superscriptaddress]{revtex4}
\usepackage{epsfig}
\usepackage[all]{xy}
\usepackage{graphicx}

\newcommand{\be}{\begin{equation}}
\newcommand{\ee}{\end{equation}}
\newcommand{\ba}{\begin{eqnarray}}
\newcommand{\ea}{\end{eqnarray}}
\newcommand{\bea}{\begin{eqnarray}}
\newcommand{\eea}{\end{eqnarray}}
\newcommand{\bean}{\begin{eqnarray*}}
\newcommand{\eean}{\end{eqnarray*}}
\newcommand{\bml}{\begin{mathletters}}
\newcommand{\eml}{\end{mathletters}}

\newcommand{\qed}{\nobreak \ifvmode \relax \else
      \ifdim\lastskip<1.5em \hskip-\lastskip
      \hskip1.5em plus0em minus0.5em \fi \nobreak
      \vrule height0.75em width0.5em depth0.25em\fi}

\begin{document}

\title{Birkhoff's theorem in Lovelock gravity}
\date{\today}
\author{Robin Zegers}
\affiliation{LPT, Universit\'e de Paris-Sud, B\^at. 
210, 91405 Orsay CEDEX, France}
\affiliation{APC, 11 place Marcelin Berthelot, F-75231 Paris Cedex 05
\footnote{UMR 7164 (CNRS, Universit\'e Paris 7, CEA, Observatoire de Paris)}}
\email{robin.zegers@th.u-psud.fr}
\preprint{LPT-05-10}
\begin{abstract}
We show that the generic solutions of the Lovelock equations with spherical,
planar or hyperbolic symmetry are locally isometric to the corresponding
static Lovelock black hole. As a consequence, these solutions are
locally static: they admit an additional Killing vector that can either be
space-like or time-like, depending on the position. This result also holds in
the presence of an abelian gauge field, in which case the solutions are
locally isometric to a charged static black hole.
\end{abstract}

\maketitle

In four dimensional General Relativity, it is well known that the spherically
symmetric solutions of Einstein's field equations in the vacuum
are locally isometric to the Schwarzschild solution --- this is Birkhoff's
theorem, see for example \cite{Straumann}. As a
consequence, these solutions are locally static outside the horizon and a spherically symmetric source does not radiate gravitational waves. In
this letter, we extend Birkhoff's theorem to Lovelock gravity. 

Lovelock theory is the most general classical theory of gravity leading to
second order field equations and conserved energy-momentum, in $D$
dimensions. The corresponding field equations without matter sources read \cite{Lovelock}
\be
\label{love}
\sum_{k=0}^{\left[ \frac{D-1}{2} \right ]} \alpha_k {\mathcal E}_{(k)a} = 0 \, ,\ee
where the brackets stand for the integer part, the $\alpha_k$ are real
constants and the ${\mathcal E}_{(k)a}$ (with $a=1 \dots D$) are given by 
\be
{\mathcal E}_{(k)a} = \left (\bigwedge_{l=1}^k \Omega^{a_{2l-1} a_{2l}} \right
) \wedge \theta^\star_{a a_1 \cdots a_{2k}} \, ,
\ee
which is of order $k$ in the curvature 2-form
$\Omega^a{}_b=\frac{1}{2}R^a{}_{bcd} \theta^c \wedge \theta^d$. Finally,
$\theta^\star_{a_1 \cdots a_k} $ is the Hodge dual of $\theta^{a_1} \wedge
\cdots \wedge \theta^{a_k}$, the basis of the space of $k$-forms
$\Omega^{(k)}(TM)$, and we thus have
\be
\theta^\star_{a_1 \cdots a_k} = \frac{1}{(D-k)!} \epsilon_{a_1 \cdots a_k
  a_{k+1} \cdots a_D} \theta^{a_{k+1}} \wedge \cdots \wedge \theta^{a_D} \, .
\ee
When $D=4$, equation (\ref{love}) reduces to Einstein equations
($k=1$) with a cosmological constant $\alpha_0$, whilst for $D=5$, the
Gauss-Bonnet term ($k=2$) must be added. For arbitrary $D$, the static
spherically symmetric solutions of (\ref{love}) were found in
\cite{Wheeler} and their extension to planar and hyperbolic symmetry is
given in \cite{Cai}. All these solutions belong to a one parameter family and read
\be
\label{ans}
g=-h(r)dt^2 + \frac{dr^2}{h(r)} + r^2 \bar{g}_{(D-2,K)} \, ,
\ee
where $h$ is given as a root of a polynomial that depends on the Lovelock
coupling constants $\alpha_k$ and on the mass parameter $\mu$, and $\bar{g}_{(D-2,K)}$ is the $(D-2)$-dimensional metric with spherical
($K=1$), planar ($K=0$) or hyperbolic ($K=-1$) symmetry and hence with isometry
groups $SO(D-1)$, $E_{D-2}$ or $SO(1,D-2)$ respectively. Notice that, in
addition to these  isometries, the solutions (\ref{ans}) also admit
$\partial_t$ as a Killing vector and this can either be space-like when $h<0$
or time-like when $h>0$, so that $g$ is locally static. This family 
contains static black holes, with one or more horizons \cite{Myers}. These
horizons can be spherical as in the  Schwarzschild case, but also planar or
hyperbolic, yielding a much richer topology as for topological black holes
\cite{Mann}. 
Though solutions of the form
(\ref{ans}) do not only describe static black holes, we shall refer to them as
static  Lovelock black hole solutions. Now, we shall prove the following
\vskip 0.25 cm
\noindent {\bf \normalsize  Theorem.} {\it The ${\mathcal C}^2$-solutions of
  the generic Lovelock field equations without matter (\ref{love}), with spherical, planar or hyperbolic symmetry are locally isometric to the
  corresponding static Lovelock  black hole solutions (\ref{ans}). }
\vskip 0.25cm
\noindent Before proceeding with the proof, note that by generic we mean that
  the Lovelock couplings $\alpha_k$ are arbitrary and independent. Indeed, as
  we shall see, when some fine-tuning conditions hold between the $\alpha_k$,
  the above theorem can be evaded. In the generic case, the theorem implies
  that the solutions of
  (\ref{love}), with spherical, planar or hyperbolic symmetry are also
  locally static; by which we mean
  that they have an additional Killing vector that is locally time-like. In
  the specific case of Einstein-Gauss-Bonnet gravity, this theorem  was
  already proven in \cite{Charmousis}. 

{\it \normalsize Proof.} We begin with a $D$-dimensional space-time with
spherical, planar or hyperbolic symmetry. We thus consider a
Lorentzian manifold $(M,g)$ admitting respectively $SO(D-1)$, $E_{D-2}$ or
$SO(1,D-2)$ as an isometry group with $(D-2)$-dimensional spacelike orbits
$\Sigma$. For all point $P\in M$, if $\Sigma_P$ is the orbit of $P$, the
tangent space $T_P M$ can be decomposed into $T_P M=T_P\Sigma_P \oplus
(T_P\Sigma_P)^\perp $. Then, let $\Sigma_P^\perp$ be the set of all the
geodesics passing through $P$ that are tangent to
$(T_P\Sigma_P)^\perp$. Locally, $\Sigma_P^\perp$ is a 2-dimensional
submanifold of $M$ that is perpendicular to the orbit $\Sigma_P$. 
Thus, on taking
$(\partial_t,\partial_z)$ as a coordinate basis of $T\Sigma^\perp$ and making
use of the conformal flatness of the 2-dimensional submanifold $\Sigma^\perp$,
one can write
\be
g=A^2(t,z)(-dt^2 + dz^2) + R^2(t,z) \bar{g}_{(D-2,K)} \, ,
\ee
where $\bar{g}_{(D-2,K)}$ is the metric over the orbits of the corresponding
isometry group. Since these orbits are invariant under their isometry group, they are
homogeneous and have constant induced curvature
\be
\bar{\Omega}^i{}_j = K \bar{\theta}^i \wedge \bar{\theta}_j \, ,
\ee 
where $i,j=1 \dots D-2$, $\bar{\theta}^i$ is the orthonormal frame adapted to
$\bar{g}_{(D-2,K)}$,
\be
\bar{g}_{(D-2,K)} = \delta_{ij} \bar{\theta}^i \otimes \bar{\theta}^j \, , 
\ee
and $\bar{\theta}^i \wedge \bar{\theta}^j$ is the resulting basis of the space 
of 2-forms on the orbits, $\Omega^{(2)}(T\Sigma) $.
It will be useful to introduce the conformal coordinates
\be
u=\frac{z+t}{2} \qquad \mbox{and} \qquad v=\frac{z-t}{2} \, ,
\ee
together with the following parametrization of the metric
\be
g= e^{2\nu(u,v)} (du \otimes dv+dv \otimes du ) + B^{2}(u,v) \bar{g}_{(D-2,K)} \, .
\ee
We define the associated orthonormal frame
\ba
\theta^u &=& e^{\nu(u,v)} du \\
\theta^v &=& e^{\nu(u,v)} dv \\
\theta^i &=& B(u,v) \bar{\theta}^i \, , 
\ea
so that
\be
g=2\theta^{(u} \otimes \theta^{v)} + \delta_{ij} \theta^i \otimes \theta^j \, .
\ee
In this basis, indices are raised and lowered using
\be
\eta_{ab} = \left ( \begin{array}{ccc}
0&1&0 \\
1&0&0 \\
0&0&\delta_{ij} 
\end{array} \right ) \, .
\ee
Given a torsionfree metric connection $\nabla$, one can define a connection
1-form $\omega^a{}_b$, such that $\omega_{ab} = -\omega_{ba}$ and $d\theta^a=
- \omega^a{}_b \wedge \theta^b$. From the latter, it is straightforward to
derive the components of the connection 1-form. 
The curvature 2-form then follows from 
\be
\Omega^a{}_b = d\omega^a{}_b + \omega^a{}_c \wedge \omega^c{}_b \, ,
\ee
yielding
\ba
\Omega^u{}_u &=& -\Omega^v{}_v = \frac{2\nu_{uv}}{e^{2\nu}} \theta^v \wedge
\theta^u  \\ 
\label{omega1}
\Omega^i{}_u &=& \frac{1}{Be^{2\nu}} \left [ \left ( B_{uu} - 2B_u \nu_u \right )
\theta^u \wedge \theta^i + B_{uv} \theta^v \wedge \theta^i \right ] \\
\label{omega2}
\Omega^i{}_v &=& \frac{1}{Be^{2\nu}} \left [ B_{uv} \theta^u \wedge \theta^i +
  \left ( B_{vv}  - 2B_v \nu_v \right
)\theta^v \wedge \theta^i \right ] \\
\Omega^i{}_j &=& \bar{\Omega}^i{}_j - \frac{2 B_u B_v}{B^{2}e^{2\nu}}
\theta^i \wedge \theta^j = \frac{K-2 B_u B_ve^{-2\nu}}{B^2} \theta^i \wedge
  \theta^j \, ,  
\ea
where $f_{u(v)} = \partial_{u(v)} f$.
Now, the projection of the $u$ (resp.~$v$)
components of (\ref{love}) onto $\theta^\star_v$ (resp.~$\theta^\star_u$)
yields the integrability conditions
\ba
\label{int1}
P'\left [ \frac{K-2B_u B_v e^{-2\nu}}{B^{2}} \right ] \left (B_{uu} -
2B_u \nu_u \right ) &=& 0 \\
\label{int2}
P'\left [\frac{K-2B_u B_v e^{-2\nu}}{B^{2}} \right ] \left (B_{vv} -
2B_v \nu_v \right ) &=& 0 \, ,
\ea
where
\be
P[X]  \equiv  \sum_{k=0}^{\left[ \frac{D-1}{2} \right ]}
\frac{\alpha_k}{(D-2k-1)!}  X^{k} 
\ee
and a prime stands for a derivative with respect to the unique argument of a
function. Notice how (\ref{int1}) and
(\ref{int2}) factorize as a product of a polynomial, times the integrability conditions
one gets from pure Einstein gravity \cite{Bowcock}. 
Up to the possible vanishing of $P'$, Einstein and Lovelock gravities thus obey
the same integrability conditions and the theorem will hold. The
projection of the $u$ (resp.~$v$) component of (\ref{love}) onto $\theta^\star_u$ (resp.~$\theta^\star_v$) then yields a further equation
\be
\label{eq}
P\left [\frac{K-2B_u B_v e^{-2\nu}}{B^{2}} \right ] -
\frac{2}{(D-1)} P'\left [\frac{K-2B_u B_v e^{-2\nu}}{B^{2}} \right ] \left ( \frac{1}{Be^{2\nu}} \left (B_{uv} - 2 \frac{B_uB_v}{B} \right ) + \frac{K}{B^{2}} \right ) = 0 \, .
\ee
Finally, the $i$ component of (\ref{love}) only projects onto
$\theta^\star_i$ giving
\ba
\label{eq2}
&& (D-1)(D-2)P\left [\frac{K-2B_u B_v e^{-2\nu}}{B^{2}}  \right ] \nonumber \\
&& - \frac{2}{B e^{2\nu}}P'\left [\frac{K-2B_u B_v e^{-2\nu}}{B^{2}}  \right ]
\left ( \frac{(2D-5)}{B}(Ke^{2\nu}-2B_uB_v) + (2D-6) B_{uv} +2B\nu_{uv}  \right )  \nonumber \\
&& +\frac{4}{B^{2}e^{4\nu}} P''\left [\frac{K-2B_u B_v e^{-2\nu}}{B^{2}}
\right ] \left ( \left (B_{uv} - 2\frac{B_uB_v}{B}
 + \frac{Ke^{2\nu}}{B} \right )^2 - \left (B_{uu}
  -2B_u \nu_u \right )  \left (B_{vv} -2B_v \nu_v \right ) \right ) = 0 \, .
\ea
Equations (\ref{int1}), (\ref{int2}), (\ref{eq}) and (\ref{eq2}) form the
full set of Lovelock equations. From the integrability conditions (\ref{int1}) and (\ref{int2}), we distinguish two classes of solutions
\begin{itemize}
\item{class I for which $P'\left [\frac{K-2B_u B_v e^{-2\nu}}{B^{2}} \right ]=0$}
\item{class II for which $P'\left [\frac{K-2B_u B_v e^{-2\nu}}{B^{2}} \right ] \neq 0$}
\end{itemize} 

For class I, 
\be
\label{cond}
K  -2 B_u B_v e^{-2\nu} = \lambda B^2 \, ,
\ee
where $\lambda$ is one of the real roots of $P'$ and is thus a function of the
$\alpha_k$. On using equation (\ref{eq}), it follows that 
$P[\lambda]=0$, whilst equation (\ref{eq2}) is trivially satisfied since
(\ref{cond}) implies that
\be
\left (B_{uv} - 2\frac{B_uB_v}{B}
 + \frac{Ke^{2\nu}}{B} \right )^2 = \left (B_{uu}
  -2B_u \nu_u \right )  \left (B_{vv} -2B_v \nu_v \right ) \, .
\ee 
Under the fine-tuning condition $P[\lambda]=0$, the solutions therefore take the form
\be
g=\frac{2B_uB_v}{K-\lambda B^2} \left (du \otimes dv + dv \otimes du \right ) +
B^2 \bar{g}_{(D-2,K)} \, ,
\ee
where $B$ is an arbitrary function of its two arguments $u$ and $v$. Such
solutions were already discussed in the case of Einstein-Gauss-Bonnet gravity \cite{Charmousis}.


For class II, the integrability conditions (\ref{int1}) and (\ref{int2}) yield
\be
B(u,v)=H\left ( F(u) + G(v) \right ) \quad \mbox{and} \quad e^{2\nu(u,v)} =
H'F'G' \, ,
\ee
where $H$, $F$ and $G$ are three functions that depend only on one argument and a prime denotes the derivative of a function with respect to
its single argument. To all functions $F$ and $G$, one can associate a new set
of coordinates $\tilde{u} = F(u)$ and $\tilde{v} = G(v)$ on $\Sigma^\perp$.
Furthermore, trading the $(\tilde{u},\tilde{v})$ conformal coordinates for
time-like $\tilde{t} = \tilde{u} + \tilde{v}$ and space-like $\tilde{z} =
\tilde{u} - \tilde{v}$, the metric can be rewritten as  
\be
g= 2H'(\tilde{z}) (-d\tilde{t}^2+d\tilde{z}^2 ) +
H^{2}(\tilde{z}) \bar{g}_{(D-2,K)} \, ,
\ee
which has a timelike Killing vector $\partial_{\tilde{t}}$ in all neighbourhood
where $H'(\tilde{z})>0$ and is thus locally
static. In particular, setting $r=H(\tilde{z})$, we can put it into the
following form
\be
\label{bh}
g=-h(r)dt^2 + \frac{dr^2}{h(r)} + r^2 \bar{g}_{(D-2,K)} \, ,
\ee
where $h(r) = 2H' (\tilde{z})$ solves equation (\ref{eq}),
\be
P \left [\frac{K-h}{r^2} \right ] - \frac{1}{D-1} P'\left [\frac{K-h}{r^2}
\right ]\left (\frac{h'}{r} + \frac{2(K-h)}{r^2} \right ) = 0 \, .
\ee
The latter can be integrated, yielding 
\be
\label{eq3}
P \left [\frac{K-h}{r^2} \right ] = \frac{\mu}{r^{D-1}} \, ,
\ee
where $\mu$ is a real constant. Let $\Lambda(\alpha_0, \cdots ,
\alpha_{\left [(D-1)/2 \right ]})$ denote a real root of $P[X]$, as a function
of the Lovelock couplings, so that we can write
\be
\label{pot}
h(r) = K - r^2 \Lambda \left (\alpha_0 - \frac{\mu}{r^{D-1}}, \alpha_1, \cdots ,
\alpha_{\left [(D-1)/2 \right ]} \right ) \, .
\ee
The metric (\ref{bh}), with $h$ given by (\ref{pot}), is the static Lovelock
black hole found in \cite{Wheeler,Cai}. Class II solutions are  the only
source-free solutions with spherical, planar or hyperbolic symmetry that are
valid for all values of the $\alpha_k$.  \qed

As we shall now see, this result still holds in the presence of an abelian gauge field that is invariant under the chosen
isometry group. Such a gauge field has a 1-form potential  $A(u,v)=L(u,v)du +
M(u,v)dv$ and therefore 
\be
F=dA = \frac{M_u - L_v}{e^{2\nu}} \theta^u \wedge \theta^v \, .
\ee
This of course implies that $dF=0$. From $d\star F=0$, on the other hand, it
follows that
\be
F=\frac{Q}{B^{(D-2)}} \theta^u \wedge \theta^v \, , 
\ee
where $Q$ is a real constant. The integrability conditions (\ref{int1}) and
(\ref{int2}) are unchanged, but there is no class I solution if $Q\neq 0$. We
are thus left with class II, which is still free of any fine-tuning, and for
which equation (\ref{eq3}) becomes
\be
P \left [\frac{K-h}{r^2} \right ] = \frac{\mu}{r^{D-1}} - \frac{Q^2}{r^{2D-4}}
\ee
as it is now sourced by the stress-energy of the gauge field. In the end, the
metric is still of the form (\ref{bh}), but now, $h$ is given by
\be
h(r) =  K - r^2 \Lambda \left (\alpha_0 - \frac{\mu}{r^{D-1}} +
  \frac{Q^2}{r^{2D-4}}, \alpha_1, \cdots ,
\alpha_{\left [(D-1)/2 \right ]} \right ) \, .
\ee
This is the Lovelock analogue of the Reissner-Nordstr\o m black hole of
General  Relativity \cite{Myers,Wiltshire}. Class II solutions are the only
solutions of the Lovelock equations, coupled to a non-vanishing abelian gauge
field, with spherical, planar or hyperbolic symmetry and they are locally static. 

Though the set of Killing vectors of space-times with spherical,
planar or hyperbolic symmetry, {\it a priori} reduces to the generators of
their respective isometry groups $SO(D-1)$, $E_{D-2}$ or
$SO(1,D-2)$, we have shown that the solutions of the Lovelock equations with
these symmetries generically get an additional Killing vector that enlarges
their isometry group, so that they reduce to locally static space-times. The
static Lovelock black holes therefore span the whole set of solutions of the
generic  Lovelock equations without matter or in the presence of an abelian
gauge field, with spherical, planar or hyperbolic symmetry. 

It is a pleasure to thank Christos Charmousis and Dani\`ele Steer for helpful
suggestions and comments.

\end{document}